\documentclass[%
 aip,
 amsmath,amssymb,
 reprint,%
]{revtex4-1}

\usepackage{graphicx}% Include figure files
\usepackage{dcolumn}% Align table columns on decimal point
\usepackage{bm}% bold math
\usepackage{braket}
\usepackage{xcolor}
\usepackage[utf8]{inputenc}
\usepackage[T1]{fontenc}
\usepackage{mathptmx}
\usepackage{etoolbox}
\usepackage{xcolor}
\usepackage{hyperref}% add hypertext capabilities
\hypersetup{colorlinks=true, linkcolor=blue, urlcolor=black, filecolor=magenta, citecolor=blue}

\makeatletter
\def\@email#1#2{%
 \endgroup
 \patchcmd{\titleblock@produce}
  {\frontmatter@RRAPformat}
  {\frontmatter@RRAPformat{\produce@RRAP{*#1\href{mailto:#2}{#2}}}\frontmatter@RRAPformat}
  {}{}
}%
\makeatother
\begin{document}

\preprint{AIP/123-QED}

%\title{Circuit quantum electrodynamics with a compact superconducting resonator}
\title{Control and readout of a transmon using a compact superconducting resonator}
% Force line breaks with \\
\author{Julia Zotova}
\affiliation{Skolkovo Institute of Science and Technology, 121205 Moscow, Russia}
\affiliation{Moscow Institute of Physics and Technology, Institutskiy Pereulok 9, Dolgoprudny 141701, Russia}
\affiliation{RIKEN Center for Emergent Matter Science (CEMS), 2–1 Hirosawa, Wako, Saitama 351–0198, Japan}
\affiliation{National University of Science and Technology `MISIS', 119049, Russia}
\email{yuliya.zotova@phystech.edu}

\author{Shtefan Sanduleanu}
\affiliation{Moscow Institute of Physics and Technology, Institutskiy Pereulok 9, Dolgoprudny 141701, Russia}
\affiliation{National University of Science and Technology `MISIS', 119049, Russia}
\affiliation{Russian Quantum Center, 121205 Skolkovo, Moscow, Russia}

\author{Gleb Fedorov}
\affiliation{Moscow Institute of Physics and Technology, Institutskiy Pereulok 9, Dolgoprudny 141701, Russia}
\affiliation{Russian Quantum Center, 121205 Skolkovo, Moscow, Russia}
\affiliation{National University of Science and Technology `MISIS', 119049, Russia}

\author{Rui Wang}
\affiliation{Research Institute for Science and Technology, Tokyo University of Science, 1-3 Kagurazaka, Shinjuku-ku, Tokyo 162-8601, Japan}
\affiliation{RIKEN Center for Quantum Computing (RQC), Wako, Saitama 351-0198, Japan} 

% \author{Yu Zhou}
% \affiliation{RIKEN Center for Quantum Computing (RQC), Wako, Saitama 351-0198, Japan} 

\author{Jaw Shen Tsai}
\affiliation{Research Institute for Science and Technology, Tokyo University of Science, 1-3 Kagurazaka, Shinjuku-ku, Tokyo 162-8601, Japan}
%\affiliation{Department of Physics, Tokyo University of Science, 1–3 Kagurazaka, Shinjuku, Tokyo 162–0825, Japan}
\affiliation{RIKEN Center for Quantum Computing (RQC), Wako, Saitama 351-0198, Japan} 

\author{Oleg Astafiev}
\affiliation{Skolkovo Institute of Science and Technology, 121205 Moscow, Russia}
\affiliation{Moscow Institute of Physics and Technology, Institutskiy Pereulok 9, Dolgoprudny 141701, Russia}

\date{\today}% It is always \today, today,

\begin{abstract}
We demonstrate control and readout of a superconducting artificial atom based on a transmon qubit using a compact lumped-element resonator. The resonator consists of a parallel-plate capacitor (PPC) with a wire geometric inductor. The footprint of the resonators is about 200~$\mu$m by 200~$\mu$m, which is similar to the standard transmon size and one or two orders of magnitude more compact in the occupied area comparing to coplanar waveguide resonators. We observe coherent Rabi oscillations and obtain time-domain properties of the transmon. The work opens a door to miniaturize essential components of superconducting circuits and to further scaling up quantum systems with superconducting transmons.
\end{abstract}

\maketitle

The first experimental evidence of  strong coupling between a single photon in a cavity and a single artificial atom was observed with a Cooper-pair-box (CPB) and a coplanar waveguide resonator~\cite{wallraff2004strong}. Such resonators are now commonly used to implement quantum harmonic oscillators on chip, mostly for qubit coupling and readout purposes~\cite{walter2017rapid, xu2020probing}. Their operating frequency is determined predominantly by geometric length, since the resonator is based on standing electromagnetic (EM) waves. Considering that the wavelength of the microwave signals propagating in coplanar lines on top of silicon at 6~GHz is about 2~cm, the enclosing area of a meandering resonator is usually above $5\times 10^5~\mu$m$^2$. For moderate resonator quantities on a single substrate, such a large size compared to the qubit incurs no significant problems, but the situation changes with the scaling of quantum integrated circuits~\cite{arute2019quantum, gong2021quantum}. Additionally, distributed resonators~\cite{jiang2022building, bolgar2018quantum} possess higher order frequency modes, which may cross-couple to the other elements of the scheme, inducing additional relaxation or decoherence~\cite{houck2008controlling}; this effect is known as frequency crowding. Therefore, there is a need to shrink the size of the readout resonators, preferably making them single-mode.

To date, there have been already a few attempts to couple a transmon capacitively to a compact resonator: lumped-element resonators with geometric inductance and planar capacitance~\cite{Rigetti2018}, Hilbert-space-filling-curves~\cite{jiang2022building}, acoustic~\cite{bolgar2018quantum}, through-silicon-vias (TSV)~\cite{hazard2023characterization} and ultrathin NbN resonators~\cite{wei2023compact} were tested. The first type of compact resonator~\cite{Rigetti2018} is the lumped-element, where its inductance is geometric and its capacitance is realized between a planar strip and the ground polygons. In the second case~\cite{jiang2022building}, a better compactness of the resonator was achieved as its inductance and capacitance were realized as special space-filling 1D structures. For acoustic resonators, based on standing  surface acoustic waves (SAW), compactness is achieved because the propagation velocity of the SAW on a piezoelectric substrate is five orders of magnitude slower~\cite{aref2016quantum} than the EM wave on a silicon substrate. Next, lumped-element through-silicon via (TSV) resonators are based on compact kinetic inductance spirals and TSV capacitors and use vertical spacial dimension to reduce the planar footprint. Finally, niobium nitride resonators~\cite{wei2023compact} achieve compactness through the high kinetic inductance of the ultrathin film.

\begin{figure}
    \centering
    \includegraphics[width=1\linewidth]{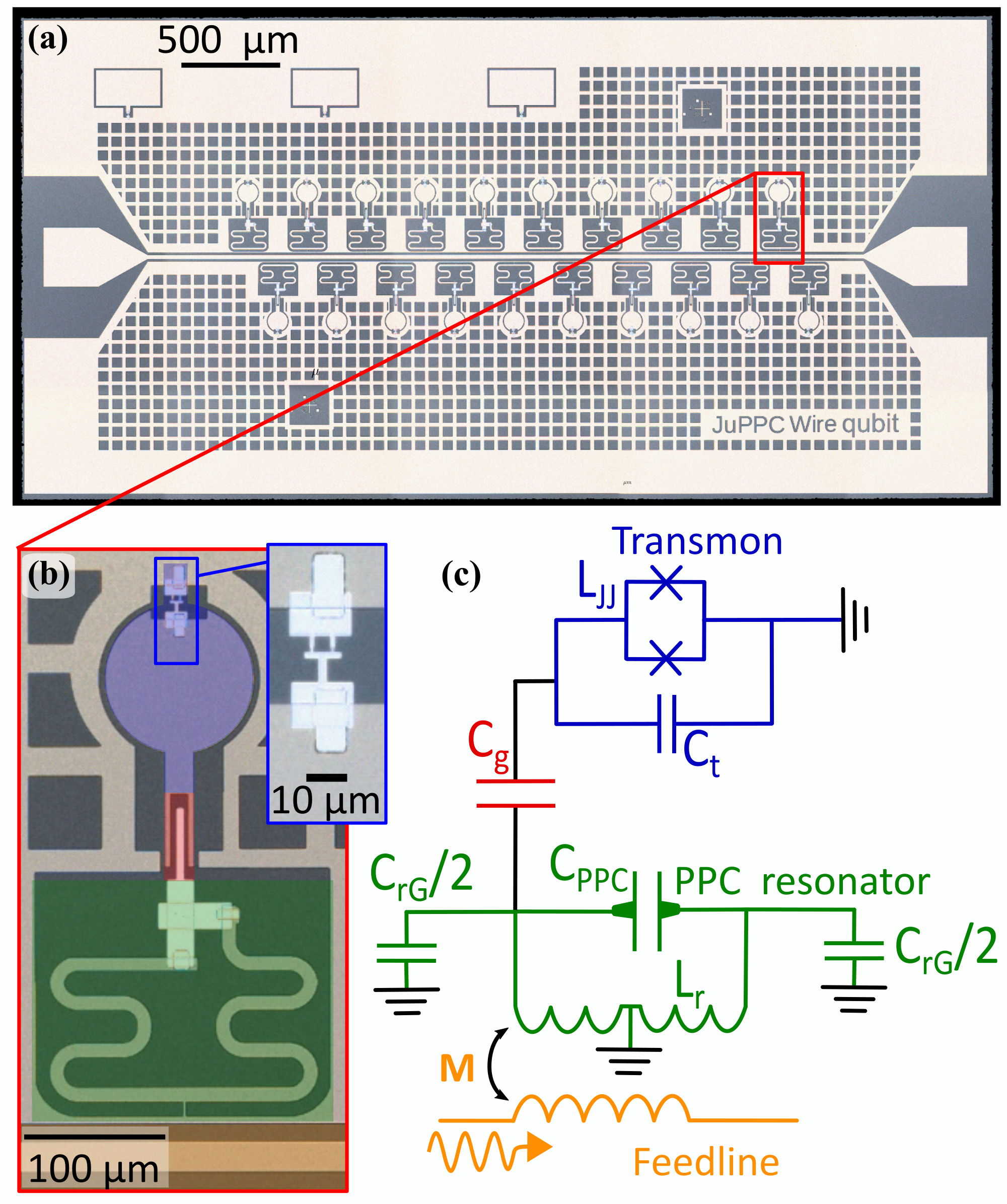}
    \caption{Investigated system and their chip.  (a) The chip image. (b): The resonator-transmon system. (c): An equivalent electric scheme of the circuit in (b).}
\label{chip}
\end{figure}

The latter three realizations allow one the best reduction of size but require special fabrication approaches, such as the use of piezoelectric substrates~\cite{bolgar2018quantum}, which usually reduce system coherence, TSV with flip-chip packaging~\cite{hazard2023characterization}, or ultrathin NbN technology~\cite{wei2023compact}. Alternative approach to realize resonators with similar sizes is based on a planar lumped-element design~\cite{zotova2023compact}.  Such parallel-plate capacitor (PPC) resonators, differently from coplanar ones, do not have higher frequency modes and are based on the standard oxidation process.  In this work, we experimentally develop cQED systems with such resonators and demonstrate coherent qubit control and dispersive readout. It consists of a common feedline (shown in orange in Fig.~\hyperref[chip]{\ref*{chip}(c)}) for 20 units of compact resonators (shown in green in Fig.~\hyperref[chip]{\ref*{chip}(c)}) and flux-tunable (via a superconducting interference device, SQUID) transmon qubit (shown in blue in Fig.~\hyperref[chip]{\ref*{chip}(c)}) connected by a coupling capacitor (shown in red in Fig.~\hyperref[chip]{\ref*{chip}(c)}). 

The lowest transmon  transition corresponds to a frequency of about 5.2~GHz and designed the same for all resonators. We keep the same inductor for all resonators with $L\approx 0.3~$nH, which we obtain from the the finite-element method (FEM) simulations.   We vary the resonator frequencies from 4.6~GHz up to 10.3~GHz by changing their capacitance, which sizes can be found in Tab.~\ref{table:resparam}. We divide all frequencies of the resonators into five groups: A (5 peaks), B (4 peaks), C (3 peaks), D (2 peaks), E (1 peak), and F (2 peaks) for clarity  with expected frequencies between 4 and 11~GHz. Since compact lumped-element resonators do not have high-frequency modes, we do not experience frequency overlaps in this setup. From finite-element simulations, we obtain the capacitance value in the circuit between the main components: resonator-qubit coupling capacitance $C_g = 6.5$~fF, shunting transmon capacitance  $C_t = 51$~fF (including Josephson capacitance) and the resonator-to-ground $C_{rG} = 58$~fF, as shown in the circuit schematics in Fig.~\hyperref[chip]{\ref*{chip}(c)}. Note that the internal resonator capacitance, formed by the PPC (see Tab.~\ref{table:resparam}), is one-two orders of magnitude larger than the resonator-to-ground capacitance. Still, $C_{rG}$ may produce additional parasitic voltage between the resonator and the transmon due to different island potentials. To avoid that, we add a short strip in the center of the resonator inductor, which does not affect the resonator dynamics.

The resonator-transmon system may be approximately described by a modified version of the Jaynes–Cummings Hamiltonian \cite{koch2007charge}: 
\[
H_{JC} = H_r + H_t + \hbar g(a^\dag b + a b^\dag),
\] 
where $H_r = \hbar \omega_r(a^\dag a +1/2)$ is the resonator Hamiltonian, 
\[\begin{aligned}
H_t &= \sqrt{8 E_J E_C} b^\dag b - \frac{1}{2} E_C b^\dag b b b^\dag \\
&= \hbar \omega_{ge} b^\dag b + \frac{1}{2}\hbar \alpha b^\dag b^\dag b b
\end{aligned}
\] 
is the approximate transmon Hamiltonian. Here, $a,\ b$ are bosonic annihilation operators  for the resonator and the qubit respectively, $\hbar \omega_{ge} = \sqrt{8E_J E_C} - E_C$, $\hbar \alpha \approx -E_C$. By $E_J, E_C$ we denote the Josephson and capacitive energies, and $|g\rangle$, $|e\rangle$, $|f\rangle$,  $|h\rangle$ denote the first four transmon eigenstates. The anharmonicity $\alpha$ is equal to $\omega_{ef} - \omega_{ge}$, and for the chosen capacitance $C_t$ $\alpha/2\pi \approx 340$ MHz. $E_J = E_J(\Phi_e) = E_J^\Sigma |\cos (\pi \Phi_e/\Phi_0)|$ can be tuned by the externally applied magnetic flux $\Phi_e$ (here $E_J^\Sigma$ is the combined energy of the two Josephson junctions in the transmon SQUID).

The coupling constant $g \equiv g_{ge}$ describing the capacitive coupling by the capacitance $C_g$ is calculated in the limit of large $C_r, C_t \gg C_g$ by~\cite{koch2007charge}
\begin{equation}~\label{eq:coupling}
\hbar g \approx V_t C_g V_{rms},
\end{equation} 
where $V_{rms} = \frac{1}{2} V^0_{rms} = \frac{1}{2} \sqrt{\hbar \omega_r/2C_r}$, $V^0_{rms}$ being the amplitude of the zero-point voltage fluctuations between the coupling electrode and the ground potential. The 1/2 prefactor appears due to the fact that the amplitude of the voltage fluctuations between the resonator and the ground is two times smaller than that of between the capacitor electrodes. Similarly, $V_t$ is the transmon dipole moment, which can be approximately expressed as zero-point fluctuations in the resonator with the transmon capacitance $V_t \approx \sqrt{\hbar \omega_{ge}/2C_t}$ in case of $E_J \gg E_C$~\cite{koch2007charge}. We than arrive at the final convenient form of~\eqref{eq:coupling}:
\begin{equation} \label{eq:coupling2}
g \approx \frac{1}{4}\frac{C_g}{\sqrt{C_r C_t}} \sqrt{\omega_r \omega_{ge}},
\end{equation} 
which for the parameters listed above and $\omega_{ge}/2\pi = 5$ GHz gives $g/2\pi \approx 15$ MHz. In the resonant case, when detuning $\Delta_{ge} = \omega_r - \omega_{ge} = 0$, the resonance frequencies of the coupled system are $E_{\pm} = \hbar\times(\omega_r \pm g)$.

The transmon measurement is done in the dispersive non-resonant regime but the absolute detuning $|\Delta_{ge}|$ in our experiments may for certain values of $\Phi_e$ become comparable to $g$ and $\alpha$. We thus provide the expression for the dispersive shift $\chi$ near and in the straddling regime~\cite{koch2007charge}:
\begin{equation}\begin{gathered}
\chi = g^2\cdot \alpha/\Delta_{ge}\Delta_{ef},
\end{gathered}\label{eq:disp_chi}\end{equation}

where $\Delta_{ef} = \omega_r - \omega_{ef}$. The transmon and resonator frequencies in this expression are slightly modified due to the Lamb shift \cite{koch2007charge}. The dispersive shift is positive if $\omega_r \in (\omega_{ef}, \omega_{ge})$ (straddling regime) and negative otherwise; close to $\omega_r = {\omega_{ef}, \omega_{ge}}$ the expression is invalid due to the brake of the dispersive regime. In the straddling regime for chosen parameters, $\chi \gtrsim 2$ MHz.

To fabricate the device, a 50~nm niobium film was deposited by magnetron spattering on a 2.5~mm $\times$ 5~mm silicon chip.  After patterning coplanar waveguides, resonator inductors and transmon capacitors are formed by reactive ion etching (RIE) in $CF_4$ atmosphere. Josephson junctions are fabricated from aluminum and aluminum oxide by using e-beam patterning and deposited by Dolan technique. Parallel-plate capacitors consist of aluminum electrodes with aluminum oxide as the dielectric layer~\cite{zotova2023compact}. To ensure good galvanic contacts between the resonator PPC and the inductor, as well as between the transmon shunting capacitor and Josephson junction interfaces, bandages~\cite{dunsworth2017characterization} are used. All patterns, except the Josephson junctions, are fabricated with photolithography. Flux traps are placed to stabilize the external magnetic field used to tune the transmon frequency. 

The chip is placed in the printed-circuit-board sample holder and then in a Cryoperm magnetic shield to protect the circuit from varying stray magnetic fields, and cooled down to 15~mK in a dilution refrigerator. The probe signals are up-converted from MHz to GHz frequencies and then attenuated to suppress thermal noise. Two synchronized continuous wave sources as local oscillators (SignalCore SC5502A), a four-channel arbitrary waveform generator (Keysight M3202A) and a four-channel digitizer (Spectrum m4x) are used for both spectroscopic and time-domain measurements. At the output, the signal is amplified by HEMT and room-temperature amplifiers and then downconverted for digitization.  The applied magnetic field is generated by applying DC-current to an inductive coil wrapped around the sample holder.

\begin{figure}
    \centering
    \includegraphics[width=1\linewidth]{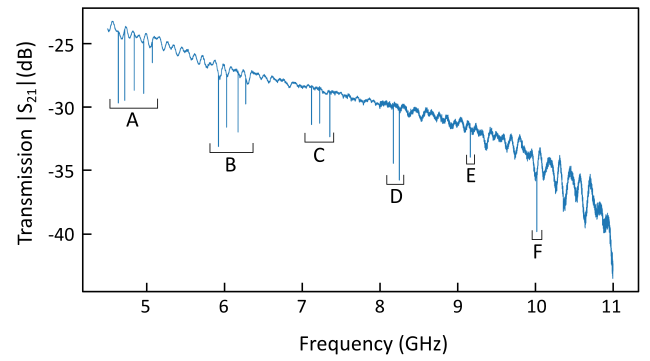}
    \caption{Amplitude of the transmission coefficient $|S_{21}|$ at $\sim$~-80~dBm on-chip power. The boxes labeled "A", "B", "C", "D", "E", "F" indicate groups with different numbers of resonators, see Tab.~\ref{table:resparam}.
    }
\label{s21}
\end{figure}

First, we obtain a plot of preliminary complex transmission coefficient $S_{21}$ at high microwave power, whose amplitude is shown in Fig.~\ref{s21}. We can clearly distinguish several groups having 5, 4, 3, 2, and 1 resonator peak(s) each according to the device design. In this measurement, the bandwidth of the HEMT amplifier was $4 - 8$~GHz, so resonator peaks above 11~GHz (18-20) are hard to detect due to poor transmission. The measured capacitance of the resonators 16F and 17F turned out to be very close to each other (see Table \ref{table:resparam}), which resulted in an overlap of the resonance peaks. Absence of any higher modes of the compact resonators confirms the lumped-element geometry. Power-dependent spectroscopy of the resonators reveals the relatively low internal quality factor $Q_\text{in} \sim 10^3-10^4$ at the single-photon limit compared to coplanar waveguide resonators, which is consistent with previous measurements of the same type of resonators~\cite{zotova2023compact}.

\begin{figure}
    \centering
    \includegraphics[width=1\linewidth]{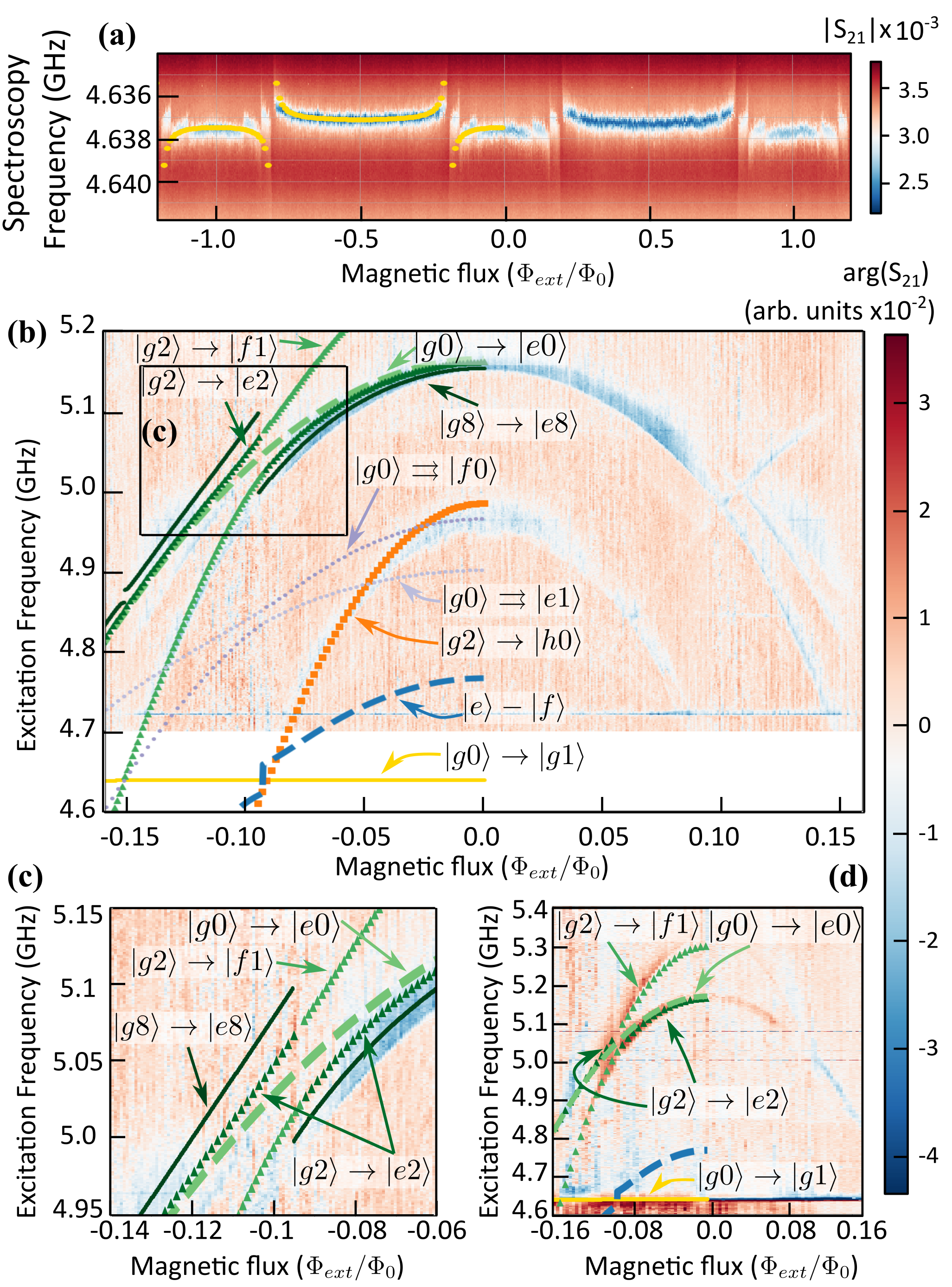}
    \caption{Experimental spectra of (a) the readout resonator of the transmon 1 A (b)-(d) transmon transitions as a function of the normalized magnetic field for the first resonator-transmon pair. (c) Enlarged part of the (b), (d) qubit spectra with higher excitation power. Fit indicates different transitions, marked with different styles and colors.}
\label{spec}
\end{figure}

Then, we record transmission spectra of the resonators as a function of the external magnetic field. As discussed above, the magnetic flux through a SQUID loop tunes the transmon frequency by modulating $E_J$.  A conventional indication of a resonator-transmon coupling is the periodic appearance of the avoided crossing, see Fig.~\hyperref[spec]{\ref*{spec}(a)}, which marks the strong coupling regime $\gamma, \kappa \ll g \ll \omega_r, \omega_{ge}$ near degeneracy points ($\Delta_{ge} = 0$)~\cite{fedorov2019automated}. Outside the degeneracy points,  the dispersive shift is significant, which is notably larger for $\omega_{ge} > \omega_{r}$ than for  $\omega_{ge} < \omega_{r}$ (for the same absolute detuning)~\cite{koch2007charge}. This is so, first, due to the increased frequency-dependent coupling constant \eqref{eq:coupling2} in the former case, and second, due to the asymmetry  of $\chi$ with reference to $\omega_r$ caused by the pole $\Delta_{ef} = 0$ of \eqref{eq:disp_chi} occurring when $\omega_{ge} > \omega_r$. This is why for the time-domain readout we only test resonators with frequencies $\omega_{ge}>\omega_{r}$. We notice that the resonator frequency on the single-tone spectra slightly fluctuates on magnetic field. However, this behavior is probably linked to TLS in the capacitor coupled to a resonator~\cite{brehm2017transmission}.\\

\begin{table*}
\small
\begin{tabular}{|p{1.5cm}|*{17}{p{.75cm}|}}
    \hline
    N, group & 1 A & 2 A & 3 A & 4 A & 5 A & 6 B & 7 B & 8 B & 9 B & 10 C & 11 C & 12 C & 13 D & 14 D & 15 E & 16 F & 17 F \\
    \hline
    $f_r$ (GHz) & 4.639 & 4.721 & 4.842 & 4.965 & 5.077 & 5.926 & 6.031 & 6.178 & 6.277 & 7.120 & 7.225 & 7.357 & 8.169 & 8.249 & 9.200 & 10.00 & 10.01 \\
    \hline
    $\sqrt{\mbox{S}}$ ($\mu$m) & 19.14 & 18.56 & 18.45 & 18.11 & 17.32 & 14.83 & 14.61 & 14.08 & 13.86 & 12.1 & 11.84 & 11.9 & 10.58 & 10.22 & 9.42 & 8.33 & 8.32 \\
    \hline
    $C_{PPC}$ (pF) & 5.13 & 4.82 & 4.76 & 4.59 & 4.2 & 3.08 & 2.99 & 2.77 & 2.69 & 2.05 & 1.96 & 1.98 & 1.56 & 1.46 & 1.24 & 0.97 & 0.97 \\
    \hline
\end{tabular}
\caption{Measured parameters of the resonators: measured resonator frequency $f_r$, linear size of capacitance area $\sqrt{S}$, fabricated capacitance $C=c \times S$, $c~\cite{zotova2023compact}\approx 14~$fF/$\mu$m$^2$.}
\label{table:resparam}
\end{table*}

For better understanding of the system energy structure vs. external magnetic field, we perform cross-Kerr two-tone spectroscopy where the first tone is set to the resonator frequency $f_r=\omega_r/(2\pi)$, and the second tone (whose frequency is on the y-axis) is aimed to excite the transitions between the eigenstates, see Fig.~\hyperref[spec]{\ref*{spec}(b-d)}. To identify the obtained transitions, we fit the two-tone spectra by numerically solving the Hamiltonian of the resonator-qubit system. The  transitions we found are labeled in Fig.~\hyperref[spec]{\ref*{spec}(b-d)}. From the spectroscopic measurement we extract the main transmon parameters. For transmon 1 A, the frequency of the main transition $\omega_{ge}(\Phi_e = 0)/2\pi = 5.19$~GHz, the anharmonicity $\alpha/h = -334~$MHz. The charging energy is thus $E_C = 334~$MHz, and $E_J^\Sigma  \approx 11.4~$GHz. The resonator-transmon coupling strength is $g/2\pi \approx 15~$MHz near 5~GHz, which is close to the value estimated by Eq.~\eqref{eq:coupling}.

\begin{figure}
    \centering
    \includegraphics[width=1\linewidth]{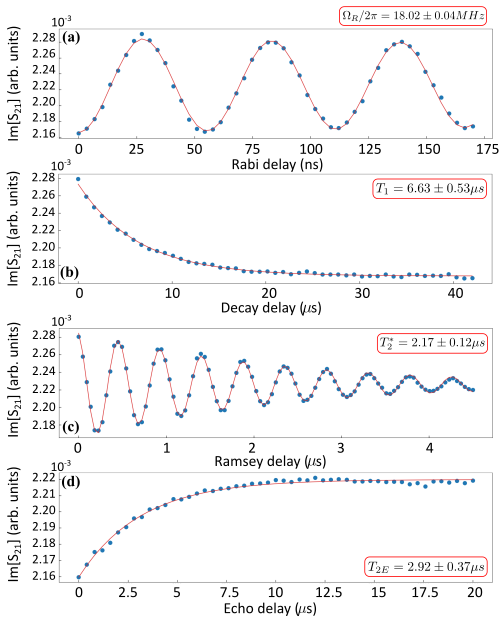}
    \caption{Experimental results of the readout of the transmon states for the first qubit. \textbf{(a)} Rabi oscillations. \textbf{(b)} The energy relaxation curve.} \textbf{(c)} The Ramsey dephasing measurements.  \textbf{(d)} The Hahn-Echo dephasing measurements.
\label{timedomain}
\end{figure}

%\autoref{spec}
In Fig.~\hyperref[spec]{\ref*{spec}(b, d)} we find the so-called "straddling regime"~\cite{koch2007charge} of the cQED system, when the readout resonator frequency $\ket{g0}-\ket{g1}$ (dashed blue line in  Fig.~\hyperref[spec]{\ref*{spec}(b, d)}) lies between the $\ket{g} - \ket{e}$ (dashed green line) and $\ket{e}-\ket{f}$ (dashed yellow line) transitions. In this regime the transition $\ket{g}-\ket{f}$ becomes allowed and we observe it spectroscopically, see Fig.~\hyperref[spec]{\ref*{spec}(d)}, $\ket{g2}-\ket{f1}$ transition (light green triangles). Note that this transition is realized with an additional photon from the resonator in our system. Additionally, we observe that the dispersive shift (or, alternatively, the Stark shift) changes its sign at the point $\Delta_{ef}=0$. For convenience, we plot numerical results for Stark-shifted main transmon transitions for $\langle n \rangle$ = 0, 2, 8  photons in the resonator.
Also, in Fig.~\hyperref[spec]{\ref*{spec}(b)} we do not observe two-photon transitions $\ket{g0} \rightrightarrows \ket{e1}$ and $\ket{g0} \rightrightarrows \ket{f0}$, which imply that we are not in the strong-pumping regime for the transmon. However, we see a surprisingly bright three-photon transition from the ground to the third excited state $\ket{h}$ of the transmon, which takes two photons from the resonator $\ket{g2} - \ket{h0}$ and manifests itself at $\omega_{gh} - 2\omega_{r}$. Probably, this high-level transition becomes allowed due to some resonance effect between the resonator and  close-tuned qubit, $\Delta_{ge} < 550~$MHz.

Then, we obtain coherent Rabi oscillations to demonstrate the quantum state control, see Fig.~\hyperref[timedomain]{\ref*{timedomain}(a)}.  The conventional pump-probe technique is used to read out the transmons. We obtain the population of the first excited state of a transmon as a function of the excitation pulse length. We extract the length of the $\pi$-pulse for the transmon excitation to the first exited state.  Then we measure the relaxation time $T_1 = 6.63 \pm 0.53~\mu s$ (Fig.~\hyperref[timedomain]{\ref*{timedomain}(b)}) by varying the delay between the $\pi$-pulse and the readout pulse.  Ramsey dephasing time $T^*_{2R} = 2.17 \pm 0.12~\mu s$ (Fig.~\hyperref[timedomain]{\ref*{timedomain}(c)}) are obtained by varying the delay between two $\pi/2$-pulses with following readout. The Hahn-Echo dephasing time $T_{2E} = 2.92 \pm 0.37~\mu s$ (Fig.~\hyperref[timedomain]{\ref*{timedomain}(d)}) was obtained by varying the delay between $\pi$-pulse and two $\pi/2$-pulses with subsequent  readout. In spite of a relatively low internal quality factor of the compact resonators $Q_\text{in} \sim 10^4$, which limit a loaded $Q_\text{loaded} \sim 10^4$  quality factor, the estimated Purcell effect tends to be small: the qubit decay limit is $T_\text{Purcell}\approx 380~\mu$s.

In summary, we demonstrate the readout of the transmon states using a compact, lumped-element resonator. The footprint of the used resonators is comparable to a transmon size ($\sim 0.04~$ mm$^2$) and more than an order of magnitude more compact than a conventional coplanar waveguide resonator ($\sim 1~$mm$^2$)~\cite{zotova2023compact}. The readout is achieved despite the relatively low internal quality factor at low photon numbers ($Q_\text{in} \sim 10^3 - 10^4$)~\cite{zotova2023compact}. In the future, the resonators can be used to to realize two-qubit gate operations. 

\begin{acknowledgments}
J.Z. thanks Yu Zhou for stimulating discussions. This work was funded by RIKEN IPA Program, by ImPACT Program of Council for Science, Technology and Innovation (Cabinet Office, Government of Japan), by CREST, JST (Grant No. JPMJCR1676) and by the New Energy and Industrial Technology Development Organization (NEDO), JPNP16007. J.Z., S.S., G.F. and O.A. also thank the support of the Russian Science Foundation, Grant No. 21-72-30026. 

\end{acknowledgments}

\section*{Data Availability Statement}

The data that support the findings of this study are available from the corresponding author upon reasonable request.

\appendix

%\section{Appendixes}

\nocite{*}
\bibliography{aipsamp}% Produces the bibliography via BibTeX.

\end{document}